\documentstyle[epsfig]{aipproc}

\def\Ls{L_{s}^{\rm obs}}
\def\Lh{L_{h}^{\rm obs}}
\def\Lsintr{L_s^{\rm intr}}

\begin{document}
\title{Physical Characteristics of the Spectral States of 
Galactic Black Holes} 

\author{Juri Poutanen$^*$, Julian H. Krolik$^{\dagger}$ and 
Felix Ryde$^{\ddag}$}
\address{$^*$Uppsala Observatory, Box 515, SE-75120 Uppsala, Sweden\\
$^{\dagger}$Johns Hopkins University, Baltimore, MD 21218 \\
$^{\ddag}$Stockholm Observatory, SE-13336 Salts\"obaden, Sweden} 

\lefthead{Spectral states of galactic black holes}
\righthead{J.~Poutanen et al.} 
\maketitle

\begin{abstract}
Using simple analytical estimates we show how the physical parameters 
characterizing different spectral states of the 
galactic black hole candidates can be  determined 
using  spectral data presently available. 
\end{abstract}

\section*{Spectral States of GBHC} 

Galactic black hole candidates (GBHC) radiate in one of several spectral 
states, and some of them switch suddenly from one state to another. 
These states can be typified by their extremes: 
a ``hard" state (HS, also called ``low", because of relatively low 
flux in standard X-ray 2 -- 10 keV band), and a ``soft" state 
(SS, a  ``high" state with relatively strong 2 -- 10 keV flux). 
The broad band spectra in both states can be described as the sum of 
a blackbody and a power-law with an exponential cut-off. 
The black body component (probably from the optically thick accretion disk) 
is more prominent in the SS, when it has 
a temperature of 0.3 -- 1~keV. The lower temperature of the 
black body in the HS makes it  difficult to detect, due to the 
interstellar absorption. The power-law energy index, $\alpha$, is 
1.0 -- 1.5 in the SS, and roughly 0.3 -- 0.7 in the HS
\cite{tanaka95,ebis96}. 
Recent OSSE observations 
reveal that the cut-off energy, $E_c$, of the power-law is correlated with 
the spectral state; the power-law 
turns over at $\sim$ 100 keV in the HS, 
and $E_c\gtrsim 200$ keV in the SS \cite{phlips96,grove97}.  
There are also indications of that the amplitude 
of the Compton reflection ``bump" increases when spectrum steepens 
\cite{ebis96}. 
 
The physical nature of the existence of the different spectral states 
and spectral transitions 
is not yet completely understood, although a number of models 
has been proposed (e.g. \cite{ichimaru77,chen96,ebisawa96,esin97}). 
Recent progress on the theoretical side (we now
understand much better how thermal Comptonization works, when the seed
photons are produced mainly by reprocessing a part of the hard
X-ray output, \cite{gh94,pk95,stern95}) and the existence of  
broad band simultaneous spectral data for some of the sources 
(e.g., \cite{grove97,gier97,zdz97}) give us an opportunity 
to use the observed spectral characteristics in these states to
determine the geometry and energy dissipation distribution in an
accreting black hole system. 
The goal of the present investigation is to infer physical parameters 
of GBHC purely on the basis of {\it radiation} physics, this later 
can be used to guide efforts to obtain {\it dynamical} explanations for
the changes in spectral state. 
A more detailed discussion can be found in \cite{pkr97}.  

\section*{Analytical Arguments} 

It is natural to attribute the two components with which the spectra
 are fitted to
two physically related regions: an optically thick (quasi-thermal) 
accretion disk, 
which is responsible for the blackbody component, and an optically thin 
hot region (corona), which radiates the hard X-rays. 
The intrinsic dissipation rates in the ``disk" and ``corona" are 
$\Lsintr$ and $L_h$, respectively. The size of the region over 
 which the ``disk" radiates most of its energy is $R_s$, and the size of the 
corona is $R_h$. 

The hard X-rays are assumed to be produced by thermal Comptonization 
(e.g. \cite{sle76})
of the seed photons that are partly created locally (by thermal bremsstrahlung
or cyclo-synchrotron radiation \cite{ny95}) and partly 
in the quasi-thermal region. The ``soft" luminosity, $L_s$ is 
partly due to local energy dissipation and partly due to reradiation 
of hard X-rays, created in the ``corona".

The shape of the Comptonized spectrum produced
by the ``corona" may be described by two parameters: 
the power-law slope $\alpha$, and the exponential cut-off energy $E_c$. 
Also two parameters (the effective temperature $T_s$ and $\Ls$) 
define the soft  part of the radiation. The relative ratio 
of the observed hard luminosity to the observed soft luminosity, $\Lh/\Ls$, 
and the magnitude of the reflection bump (described by the 
parameter, $C$, the fraction of solid angle that the  optically thick region 
occupies around the ``corona") complete the set of observables.

These phenomenological parameters are determined
by two dimensional quantities, the total dissipation rate and $R_s$, and 
four dimensionless physical parameters: the ratio $\Lsintr/L_h$;
the Compton optical depth of the ``corona" $\tau_T$; the fraction $D$ of the 
light emitted by the thermal region
which passes through the ``corona"; and the ratio $S$ of intrinsic seed photon
production in the ``corona" to the seed photon luminosity injected from outside.
Another dimensionless parameter, the compactness
$l_h \equiv L_h\sigma_T/(m_e c^3 R_h)$, may be used to determine
the relative importance of $e^{\pm}$ pairs in the corona.  In this
context it is also useful to distinguish the net lepton Compton optical depth
$\tau_p$ from the total Compton optical depth (including pairs), $\tau_T$.

   We will show how all these parameters, as well as several others of
physical interest, may be inferred from observable quantities.  

   Some of the physical parameters of the system may be derived (or
at least constrained) almost directly from observables.  For example,
the electron temperature in the corona (measured in electron rest mass
units) is very closely related to the cut-off
energy of the hard component: 
\begin{equation}  
\Theta \simeq f_x E_c/(m_e c^2) \, , 
\end{equation} 
where $f_x \sim 0.7$. Similarly,  
\begin{equation}
L_h \simeq  \Lh/\{ 1 - C [ 1 - a ]\},
\end{equation}
where $a$ is the albedo for Compton reflection. 
The intrinsic disk luminosity is
\begin{equation}
\Lsintr \simeq  \Ls - C L_h [ 1 - a ] . 
\end{equation}
Taking the local disk emission to be approximately black body,
disk's inner radius is
\begin{equation}
R_s \simeq \left\{ \Ls / 4\pi \sigma T_{s}^4 \right\}^{1/2} , 
\end{equation}
where $T_s$ is the effective temperature at the inner edge. 
A number of correction factors (accounting for difference between 
color and local effective temperature \cite{st95} and 
incorporating the general relativistic corrections \cite{zcc97} etc.)
were omitted in these formulae. 

 We next employ the two following analytic scaling approximations for
thermal Comptonization spectra found by \cite{pk95}: 
\begin{equation}
D(1 + S) = 0.15 \alpha^4 L_h / [\Lsintr + CL_h (1 - a)] 
\end{equation}
and
\begin{equation}
\tau_T = 0.16 /(\alpha \Theta) .
\end{equation}
The first expresses how the power-law hardens as the heating rate in the
corona increases relative to the seed photon luminosity; the second
expresses the trade-off in cooling power between increasing optical depth
and increasing temperature.  

  Finally, both $C$ and $D$ may be written in terms of $R_h$ and $R_s$.  If
the corona is a sphere centered on the black hole, and the disk is an
annulus of inner radius $R_s$ and infinitesimal vertical thickness,
\begin{equation}
\label{eq:cd}
C \simeq R_h/(6R_s) \quad \mbox{and} \quad D  \simeq (1/16)(R_h/R_s)^3 . 
\end{equation}
The coefficient in the definition of $C$ is most accurate when the emissivity
is uniform within the sphere and it has order unity optical depth; the
coefficient in the definition of $D$ assumes the disk surface brightness
is $\propto r^{-3}$ and its intensity distribution is isotropic. 
 The relationship between $R_h/R_s$ and $C$ or $D$ is somewhat different
in other geometries, but the scaling tends to be similar.
Equations (\ref{eq:cd}) 
may be regarded as giving {\it independent}
estimates of $R_h/R_s$.  If $C$ is used, this ratio is constrained by the
observed magnitude of the Compton reflection bump; if $D$ is used, it is
determined by the observed hard and soft luminosities, and the slope of
the hard X-ray power-law. The corona is assumed roughly spherical.

\section*{Results and Conclusions} 

The method described above  can be applied to find 
physical parameters of the system $\Theta, \tau_T, R_s, D(1+S)$, 
$R_h/R_s, \Lsintr/\Lh$ and $l_h$ from the set of observables 
$\alpha, E_c, T_s, \Ls, C$ and $\Lh/\Ls$. 
Unfortunately, existing data are good enough only for a small number 
of objects. This method applied to Cyg X-1 (see \cite{pkr97}) 
shows that the inner edge of the cold disk shrinks by roughly a factor of 5 
between the hard and soft states (from $\sim 40 r_g$ to $\sim 8 r_g$, 
assuming 10 $M_{\odot}$ black hole), while the corona shrinks in radial 
scale by only factor of two. In the SS, corona  covers a sizable portion 
of the inner disk (see Fig.~1 and \cite{esin97,gier97}). 

\begin{figure}[b!] 
\centerline{\epsfig{file=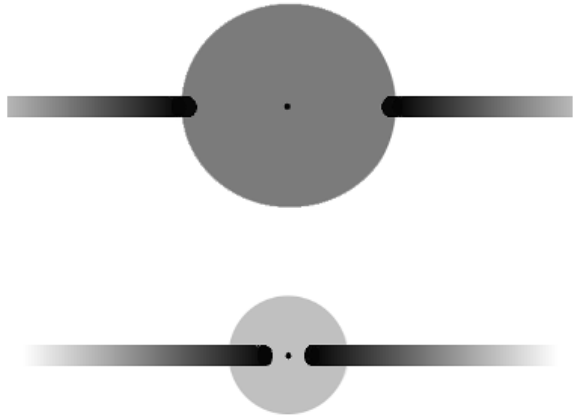,height=2.5in,width=3.5in}}
\vspace{10pt}
\caption{Geometry of the accretion flows around black holes in
the hard (top) and soft (bottom) states.
} 
\label{fig1}
\end{figure}

Despite the change in coronal size and luminosity, the compactness of
the corona almost does not change during the hard-soft transition. 
However, its optical depth drops by an order of magnitude from 
$\sim 1-2$ to $\sim 0.1$. Electron temperature of the corona is 
$\sim 100$ keV in the HS and probably is higher in the SS.  
In the HS the disk receives only a minority of the dissipation, 
$\Lsintr/L_h \sim 0.1$, but  is the site of most of the heat release in the 
SS, $\Lsintr/L_h \sim 3$. 

Detailed calculations based on the method by \cite{ps96}, that allow us 
to solve for the energy and electron-positron pair balance together 
with the self-consistent treatment of the Comptonization in the corona, 
confirm the conclusions made from  simple analytical arguments. 
We also are able to show that in the case of thermal electrons 
there are very few  $e^{\pm}$ pairs in the HS, but they can be comparable 
in number to the net electrons in the SS. In contrast the amount 
of pairs in the SS  can be significant. This means that the fall 
in $\tau_p$ from the HS to the SS is greater than the fall in $\tau_T$. 

We close by noting that simultaneous broad band observations 
from soft X-rays up to gamma-rays are necessary 
in order to make strong conclusions regarding the geometry 
and physical conditions in the accretion flows around black hole.


\end{document}